\documentstyle[pra,aps,epsfig]{revtex}

\begin{document}

\bibliographystyle{prsty}

\draft

\title{Two-component Fermi gas in a one-dimensional harmonic trap}

\date{Received 28 August 2001}

\author{Gao Xianlong and W. Wonneberger}

\address{Abteilung f\"ur Mathematische Physik, Universit\"at Ulm, D89069 Ulm, Germany}
      
\maketitle

\begin{abstract}
A many body theory for a two-component system of spin polarized interacting fermions 
in a one-dimensional 
harmonic trap is developed. The model considers two different states of the same 
fermionic species and treats the dominant interactions between the two using the 
bosonization method for forward scattering. Asymptotically exact results for the 
one-particle matrix elements at zero temperature are given. Using them, occupation 
probabilities of
oscillator states are discussed. Particle and momentum densities are calculated 
and displayed. It is demonstrated how interactions modify all these quantities. An 
asymptotic connection with Luttinger liquids is suggested. The relation of the 
coupling constant of the theory to the dipole-dipole interaction is also discussed. 
\end{abstract}


\section{Introduction}
The achievement of Bose-Einstein condensation in dilute ultracold gases \cite{ADB95} 
renewed the interest in fermionic many body systems \cite{BDL98,BB98,ZGOM00}
and their superfluid properties \cite{HFSt97,BP98,HSt99,C99}.
Recent experimental successes in obtaining degeneracy in three dimensional Fermi 
vapors \cite{DMJ99,SFCCKMS01} intensified the interest in confined Fermi gases. 

Using microtrap technology \cite{VFP98,FGZ98,DCS99,RHH99}, it will become
possible in the near future to produce a neutral ultracold  
quantum gas of quasi one-dimensional degenerate fermions. 

In many cases, identical spin polarized fermions experience only 
a weak residual interaction because s-wave scattering is forbidden. 
This restriction does not hold for a two-component system of spin 
polarized fermions and significant 
interactions between the components are possible. For instance, the  
dipole-dipole interaction \cite{GER01} can become relevant, especially in the case 
of polar molecules \cite{BBC00}.

The confinement of a trapped ultracold gas can be realized by a harmonic potential,
which is more realistic than trapping between 
hard walls ("open boundary conditions"). The latter system of interacting one-dimensional 
fermions constitutes a bounded Luttinger liquid, which allows an exact treatment for certain
types of interactions \cite{C84,EA92,FG95,EG95,WVF96,MEJ97,VYG00}.

In this article, we consider a quasi one-dimensional spin polarized Fermi gas composed of 
an equal number of atoms in two different internal states and confined by a harmonic potential.
One possible realization consists of trapped (electron) spin polarized fermions in 
two different hyperfine states as discussed in \cite{HFSt97} for the case of $^6Li$.
 
We consider the inter-component interaction between the two components. As in \cite{WW01}, 
we apply the bosonization method known from the Luttinger model (for reviews we 
refer to \cite{E79,V95,Sch95}) to treat the interactions. The generalization to two 
components is analogous to the inclusion of spin $\hbar/2$ into the Luttinger model 
\cite{LE74}. The bosonization method relies on fermion-boson transmutation in one spatial 
dimension: Physical quantities can be calculated in a bosonic formulation instead of the 
fermionic theory, and the two calculations give the same answer \cite{DL73,F76,GMR79,Haldane}. 

We show how interactions modify the one-particle properties of the two-component Fermi gas.
Results for the non-interacting Fermi gas in a one-dimensional harmonic trap 
were given in \cite{VMT00,GWSZ00}.

The paper is organized as follows. Section II develops the theory for the two-component case. 
Section III applies the theory to the calculation of the one-particle matrix elements.
In Section IV, occupation probabilities, off-diagonal matrix elements, and densities of
particles and momenta are evaluated numerically for two different interaction models. Section
V discusses the problem of the Fermi edge in the present case and a possible relation to the
standard Luttinger model result is pointed out. In Section VI, the relation of the coupling 
constant of the theory to the dipole-dipole interaction is discussed. An Appendix is concerned 
with the bosonization procedure for bilinear forms of auxiliary fields in the case of two 
components.

\section{Two-Component Theory}

The two component Fermi gas of uniform mass $m_A$ is confined by the one-dimensional
harmonic potential

\begin{equation}\label{1.1}
V(z) = \frac{1}{2} m_A \omega ^2 _{\ell} z^2,
\end{equation}

with longitudinal trap frequency $\omega _\ell$. The unperturbed Hamiltonian in second 
quantization is

\begin{equation}\label{1.2}
\hat{H}_0 = \sum^\infty _{n=0,\sigma=\pm 1} \hbar \omega _n\,\hat{c}^+_{n \sigma}
 \hat{c}_{n \sigma}.
\end{equation}

The index $\sigma=\pm 1$ refers to the two components and $\hat{c}_{n \sigma}^+$ creates
a fermion of species $\sigma$ in the oscillator state $|n \rangle$.

The one-particle energies

\begin{equation}\label{1.2a}
\hbar \omega _n = \hbar \omega _{\ell} (n+1/2),\,\,\, n=0,1,...\,, 
\end{equation}

are seen to depend linearily on the quantum number $n$ of oscillator states.
This is one of the requirements for bosonization. In addition, exact solvability 
rests on the presence of an anomalous vacuum (cf. \cite{E79,V95,Sch95}), which
is constructed by extending
the linear dispersion of oscillator states to arbitrarily negative energies 
and then filling all states of negative energy. 
However, the anomalous vacuum has little effect for processes near the
Fermi energy $\epsilon _F = \hbar \omega _\ell(N-1/2) $ provided $N$ is sufficiently large
\cite{Haldane}, making the treatment asymptotically exact.

The success of the Luttinger model is based on the possibility to express forward 
scattering processes entirely in
terms of the density fluctuation operators. For the two component system, these 
operators are

\begin{eqnarray}\label{1.3}
 \hat{\rho}_\sigma (p) \equiv \sum _{q} \hat{c}^+_{q+p\, \sigma} \hat{c}_{q \sigma}. 
\end{eqnarray}

Due to the presence of the anomalous vacuum, they obey bosonic commutation relations

\begin{eqnarray}\label{1.4}
[\hat{\rho}_\sigma (-p), \hat{\rho}_{\sigma'} (q)] = p\, \delta _{\sigma, \sigma'}\,
\delta _{p,q}.
\end{eqnarray}

In our case, the interaction Hamiltonian is given by a two particle interaction 

\begin{eqnarray}\label{1.5}
\hat{V} =\frac{1}{2} \sum _{mnpq,\sigma,\sigma'}
V(m \sigma',p \sigma;q\sigma',n \sigma)\,(\hat{c}^+_{m \sigma'} \hat{c}_{q \sigma'} )
( \hat{c}^+_{p \sigma} \hat{c}_{n \sigma})
\end{eqnarray}

without "component flip", i.e., the possibility for a fermion to change its state
$\sigma$ in the collision process is excluded.

The case $\sigma = \sigma\prime$ corresponds to a weak intra-component interaction 
$V_\parallel$, while
$\sigma = -\sigma \prime$ is a relevant inter-component interaction $V_\perp$.

Similar to \cite{WW01}, two cases of solvable forward scattering processes can be identified:

\begin{equation}\label{1.6}
\hat{V} = \hat{V_a}+\hat{V_b}
\end{equation}

with

\begin{eqnarray}\label{1.7}
\hat{V_a}&=&\frac{1}{2} \sum _{p,\sigma} V_{a \parallel}(|p|)\, \hat{\rho} _{\sigma}(-p)
\,\hat{\rho}_{\sigma}(p)+\frac{1}{2} \sum _{p,\sigma} V_{a \perp}(|p|)\,\hat{\rho}_{- \sigma}(-p)
\,\hat{\rho}_{\sigma}(p),
\\[3mm]\nonumber 
\hat{V_b}&=&\frac{1}{2} \sum _{p,\sigma} V_{b \parallel}(|p|)\,\hat{\rho}_{\sigma}(p)\,
\hat{\rho}_{\sigma}(p)+\frac{1}{2} \sum _{p,\sigma} V_{b \perp}(|p|)\,\hat{\rho}_{\sigma}(p)
\hat{\rho}_{-\sigma}(p).
\end{eqnarray}

The coupling functions $V_{a \perp}$ and $V_{b \perp}$ are the analogues of $g_{4 \perp}$
and $g_{2 \perp}$ of the Luttinger model.

Forward scattering dominates when the pair interaction is sufficiently long ranged, e.g.,
in the case of dipole-dipole interactions.
In \cite{GW01}, a detailed discussion is given, how the assumed forms (\ref{1.7}) can
be related to real scattering potentials.

The case of two components requires canonical transformations to mass fluctuation 
operators

\begin{eqnarray}\label{1.8}
\hat{\rho}(p)\equiv \frac{1}{\sqrt{2}}\,\left[\hat{\rho}_+(p)+\hat{\rho}_-(p)\right],
\end{eqnarray}

and component fluctuation operators

\begin{eqnarray}\label{1.9}
\hat{\sigma}(p)\equiv\frac{1}{\sqrt{2}}\,\left[\hat{\rho}_+(p) -\hat{\rho}_-(p)\right],
\end{eqnarray}

such that the Hamiltonian for low lying excitations separates into
$\tilde{H}=\tilde{H}_\rho + \hat{H}_\sigma$ in analogy to the spin $1/2$ case \cite{LE74}
of the Luttinger model. 

The transformation to new bosonic operators

\begin{eqnarray}\label{1.10}
\hat{\rho} (p) = \left \{
\begin{array}{lll}
\sqrt{|p|} & \hat{d}_{|p|+}, & p< 0,\\
\sqrt{p}   & \hat{d}^+_{p+}, & p> 0,
\end{array} \right.\quad 
\hat{\sigma} (p) = \left \{
\begin{array}{lll}
\sqrt{|p|} & \hat{d}_{|p|-}, & p< 0,\\
\sqrt{p}   & \hat{d}^+_{p-}, & p> 0,
\end{array} \right. 
\end{eqnarray}

leads to canonical commutation relations

\begin{eqnarray}\label{1.11}
[\hat{d}_{m \mu}, \hat{d}^+_{n \nu} ] = \delta _{\mu, \nu}\,\delta _{m, n}. 
\end{eqnarray}

The new label $\nu=\pm 1$ refers to mass and component fluctuations.
 
The bosonic version of the unperturbed Hamiltonian in the N-fermion sector is 
\cite{Haldane,deK35}:
 
\begin{eqnarray}\label{1.12}
\tilde{H}_0 = \frac{\hbar \omega _\ell}{2}\, \sum _{m>0,\nu}  m \,\left\{\hat{d}^+_{m \nu}
 \hat{d}_{m \nu} +\hat{d}_{m \nu} \hat{d}^+_{m \nu}  \right\}.
\end{eqnarray}

The complete bosonic interaction operator becomes

\begin{eqnarray}\label{1.13}
 \hat{V} &=& \frac{1}{2} \sum _{m>0,\nu} m
               [V_{a\parallel} (m)+\nu \,V _{a \perp} (m) ]
               \left\{\hat{d}^+_{m \nu}
               \hat{d}_{m \nu} +\hat{d}_{m \nu} \hat{d}^+_{m \nu}  \right\}
\\[4mm]\nonumber
         &+& \frac{1}{2} \sum _{m>0,\nu} m
               [V_{b\parallel} (m)+\nu \,V _{b \perp} (m) ]
               \left\{ \hat{d}^{+2}_{m \nu} + \hat{d}^2_{m \nu} \right\}. 
\end{eqnarray}

The total bosonic Hamiltonian is diagonalized in a standard way using the Bogoliubov 
transformation

\begin{eqnarray}\label{1.13a}
 \hat{d} _{m \nu} =\hat{S}^+\hat{f}_{m \nu}\hat{S}
 = \hat{f} _{m \nu} \cosh \zeta _{m \nu} - \hat{f}^+ _{m \nu} \sinh \zeta _{m \nu}, 
\end{eqnarray}

with

\begin{eqnarray}\label{1.13b}
\hat{S}=\exp \left\{\frac{1}{2}\sum _{m>0, \nu=\pm 1}\zeta _{m \nu}(\hat{f}_{m \nu}^2
-\hat{f}^{+2}_{m \nu}) \right\}.
\end{eqnarray}

The transformation parameters $\zeta _{m \nu}$ are determined by the diagonalization 
conditions

\begin{equation}\label{1.14}
\tanh(2 \zeta _{m \nu}) = \frac{V_{b \parallel} (m)+\nu\,V_{b \perp} (m)}
                               {\hbar \omega _\ell + V_{a\parallel} (m)+\nu V_{a \perp} (m) }. 
\end{equation}

Finally, we arrive at the free bosonic Hamiltonian

\begin{eqnarray}\label{1.15}
\tilde{H}=\sum _{m>0,\nu} m\,\epsilon _{m \nu} \hat{f}^+_{m \nu} \hat{f}_{m \nu} 
+\mbox{const.}, 
\end{eqnarray}

describing density wave excitations in the two-component Fermi gas. The excitation spectra 
are

\begin{equation}\label{1.16}
\epsilon _{m \nu} = \frac{ \hbar \omega _\ell + V_{a \parallel} (m) + \nu V_{a \perp} (m)}
{\cosh(2 \zeta _{m \nu})}. 
\end{equation}

In connection with the calculation of one-particle matrix elements, scaled coupling 
constants 

\begin{eqnarray}\label{1.17}
\alpha _{m \nu} \equiv \frac{1}{2}\, \sinh(2 \zeta _{m \nu}),\quad 
\gamma _{m \nu} \equiv \sinh^2 \zeta _{m \nu}
\end{eqnarray}

will appear.

Usually, the intra-component scattering is negligible ($V_\parallel \rightarrow 0$,
the case $V_\parallel \ne 0$ was considered in \cite{WW01} for the one-component system) 
and dominant forward scattering for the inter-component part results \cite{GW01} in

\begin{eqnarray}\label{1.18}
V_{a \perp}(m)= V_{b \perp}(m) \equiv V(m)\,\hbar\,\omega _\ell.
\end{eqnarray}

Then the simpler relations

\begin{eqnarray}\label{1.19}
\epsilon _{m \nu} = \hbar \omega _\ell\,\sqrt{1+ 2 \nu\,V(m)},\quad
\alpha _{m \nu} = \frac{\nu\,V(m)}{2 \sqrt{1+ 2 \nu\,V(m)}}
\end{eqnarray}

hold for $|V(m)|<\hbar \omega _\ell/2$. 

Following \cite{WW01}, we will also consider two specific interaction models:
A simplified model IM1, when only one mode $V(m)=V(1)(\delta _{m,1}+
\delta _{m,-1})$ contributes. This model preserves many of the features of the 
interaction in the full model (interaction model 2, see IM2 below), when infinitely 
many modes are superimposed.

In the case of IM1, the relevant coupling constants are

\begin{eqnarray}\label{1.20a}
\zeta _{1 \nu} = \frac{1}{2}\,\mbox{artanh}\left(\frac{\nu\,V(1)}{1 +\nu V(1)}
\right),\quad
\alpha _{1 \nu} =  \frac{1}{2}\sinh(2 \zeta _{1 \nu}),\quad
\gamma _{1 \nu} = \frac{1}{2}\left(\sqrt{1+4\,\alpha _{1 \nu}^2}-1 \right).
\end{eqnarray}

In the case of IM2, the coupling constants decay exponentially according to

\begin{eqnarray}\label{1.20b}
\alpha _{m \nu}=\exp(-r_\alpha m/2)\,\alpha _{0 \nu},\quad
\alpha _{0 \nu} = \exp(r_\alpha/2 )\,\alpha _{1 \nu},\quad
\gamma _{m \nu} = \exp(-r_\gamma m)\,\gamma _0, \quad \gamma _{0 \nu} =\exp(r_\gamma)
\,\gamma _{1 \nu}.
\end{eqnarray}

An essential step in the actual calculation of physical quantities is the connection between
fermionic operators and bosonic fields. 
The bosonization of fermion generation and destruction operators 
is completely solved for the Luttinger model \cite{Haldane,LP74,M74,HSU80}. 

In the present case, the situation is less comfortable: Apart from the above 
association of mass and component fluctuation operators with the $\hat{d}$ operators, 
we can bosonize only bilinear forms of an auxiliary field
following the prescription of \cite{SchM96}. The auxiliary field is defined by

\begin{equation}\label{1.21}
 \hat{\psi} _{a \sigma} (v) \equiv \sum^\infty _{l=-\infty} e^{ilv} \hat{c}_{l \sigma}
= \hat{\psi}_{a \sigma} (v + 2 \pi). 
\end{equation}

The Appendix demonstrates that the required bosonization for a two-component Fermi gas is
 
\begin{eqnarray}\label{1.22}
 \hat{\psi}^+_{a \sigma} (u) \hat{\psi}_{a \sigma} (v) = G_N ( u - v )
\,\exp \left\{ - i \left( \hat{\phi}_\sigma^+(u) - \hat{\phi}_\sigma^+(v) \right) \right\}
\,\exp \left\{ - i \left(\hat{\phi}_\sigma (u) -\hat{\phi}_\sigma (v) \right) \right\},
\end{eqnarray}

using the two-component non-Hermetian bosonic field 

\begin{eqnarray}\label{1.23}
\hat{\phi}_\sigma (v) = -i \sum^\infty _{n = 1} \frac{1}{ \sqrt {2 n} }\, e^{i n v}
\,(\hat{d}_{n+} +\sigma \hat{d}_{n-})\neq \hat{\phi}_\sigma^+(v).
\end{eqnarray} 

The distribution-valued prefactor  $G_N (u)$ is the same as in \cite{SchM96}:

\begin{eqnarray}\label{1.24}
\quad G_N (u) = \sum^{N-1} _{l = - \infty} e^{-i l (u+i \eta)}.
\end{eqnarray}

\section{One-Particle Matrix Elements}

The above prescription allows to calculate analytically all m-particle matrix elements
of bilinear fermion operators. 

It is not difficult to carry the calculation of one-particle matrix
elements in \cite{WW01} over to the present case of two components:

\begin{eqnarray}\label{1.25}
 \langle \hat{c}^+ _{n \sigma}  \hat{c}_{q \sigma} \rangle = 
 \sum _{l=-\infty}^{N-1} \int^{2 \pi}_0 \int^{2 \pi}_0 \,\frac{du dv}{4 \pi^2}
e^{i(n-l)(u+i \epsilon) - i(q-l)(v-i \epsilon)}
\langle e^{-i \hat{\phi}_\sigma^+ (u) + i	\hat{\phi}_\sigma^+ (v)}
 e^{-i	\hat{\phi}_\sigma (u) + i  \hat{\phi}_\sigma (v)} \rangle.  
\end{eqnarray}
 
Using the bosonic Wick theorem, the expectation value 
$\langle \quad\rangle \equiv \exp[-W_\sigma]$ on the r.h.s. can be evaluated. At zero 
temperature, the function $W_\sigma$ is given by
 
\begin{eqnarray}\label{1.26}
 W_\sigma= W_\sigma(u,v) = \sum _\nu \sum^\infty _{m=1} \frac{1}{m} 
\left[\gamma _{m \nu} - \alpha _{m \nu} \,\cos m(u + v) \right]
\left\{ 1 - \cos m (u - v) \right\}.
\end{eqnarray}

This quantity is independent of component label $\sigma$, as expected. 

Comparing (\ref{1.26}) with (39) in \cite{WW01}, it is seen that the effective 
coupling constants in the two-component case are

\begin{eqnarray}\label{1.26a}
\bar{\alpha} _m=\frac{1}{2}\,\sum _{\nu=1}^2 \alpha _{m \nu},\quad
\bar{\gamma} _m=\frac{1}{2}\,\sum _{\nu=1}^2 \gamma _{m \nu}.
\end{eqnarray}

W is a real and an even function of its arguments leading to the symmetries

\begin{eqnarray}\label{1.27}
\langle \hat{c}^+ _{n \sigma} \hat{c} _{q \sigma} \rangle =\langle \hat{c}^+ _{q \sigma}
 \hat{c} _{n \sigma} \rangle
=\langle \hat{c}^+ _{n \sigma} \hat{c} _{q \sigma} \rangle^*
\end{eqnarray} 

and to the condition $n+q=2m$, $m=0,1,2,...\,$.

For the interaction model IM1, one of the integrations in (\ref{1.26}) can be performed 
giving the closed expression for the matrix elements of each component:

\begin{eqnarray}\label{1.28}
M(m,p)\equiv \langle \hat{c}^+_{m-p}\hat{c}_{m+p}\rangle=\frac{1}{2}\delta _{p,0}
-&&\frac{1}{2\pi}\int _{-\pi}^\pi ds\left\{\frac{\sin((m+1/2-N)s)}{2 \sin(s/2)}\right\}
\\[4mm]\nonumber
&&\times\exp[-2 \bar{\gamma} _{1}(1-\cos(s))]\,I_p(2 \bar{\alpha} _{1}(1-\cos(s))).
\end{eqnarray} 

Due to the factor $\{\sin(...)\}$, the following symmetries hold:

\begin{eqnarray}\label{1.29} 
\langle \hat{c}^+ _{2N-1-m-p \,\sigma}\hat{c} _{2N-1-m+p \,\sigma} \rangle
= \delta _{p,0}-\langle \hat{c}^+ _{m-p \,\sigma}\hat{c} _{m+p \,\sigma} \rangle.
\end{eqnarray}

Similarily, IM2 leads to:

\begin{eqnarray}\label{1.30}
M(m,p)=
\frac{1}{2}\,\delta _{p,0}
-&&\int^{\pi}_{-\pi}\frac{dt}{2 \pi}\, \frac{\cos(p \,t)}{[1+Z_\alpha-\cos(t)]
^{\bar{\alpha} _{0}}}
\,\int^{\pi}_{-\pi} \frac{ds}{2 \pi}\,\left\{\frac{\sin((m+1/2-N)s)}{2 \sin(s/2))}\right\}
\\[4mm]\nonumber
&&\times\left[\frac{Z_\gamma}{1+Z_\gamma-\cos(s)}\right]^{\bar{\gamma} _{0}}
\,[(1+Z_\alpha - \cos(t-s))(1+Z_\alpha-\cos(t+s))]^{\bar{\alpha} _{0}/2}, 
\end{eqnarray} 

with decay parameters

\begin{eqnarray}\label{1.31}
Z_\gamma=\cosh(r_\gamma)-1,\quad Z_\alpha=\cosh(r_\alpha/2)-1.
\end{eqnarray} 

\section{Numerical Results}

The main results of the paper are the formulae (\ref{1.28}) and (\ref{1.30}) for
the one-particle matrix elements. They are identical in form to those in \cite{WW01},
depend, however, differently on the coupling constants. This leads to very different 
physical predictions, which are presented in a number of figures for fermion numbers
$2N=14+14$. 

Using (\ref{1.14}) and (\ref{1.17}), it is found that the main coupling parameters
$\bar{\alpha}_m$

\begin{eqnarray}\label{2.1}
\bar{\alpha} _m =\frac{V(m)}{4}\left\{\frac{1}{\sqrt{1+2\,V(m)}}
-\frac{1}{\sqrt{1-2\,V(m)}} \right\} 
\end{eqnarray}

are non-positive and even functions of the interactions $V(m)$: Irrespective of 
the sign of the interaction between the two components, the effective interaction
in each component of the Fermi gas is attractive.  

In the case of IM1, only $\bar{\alpha}_1$ is needed as input parameter in the calculation 
of the matrix elements. $|V(1)|$ is obtained via (\ref{2.1}) and all other quantities 
such as $\zeta _{1 \nu}$ and $\bar{\gamma}_1$ can be calculated from (\ref{1.20a}) and
(\ref{1.26a}).

We start with the discussion of the occupation probabilities $P(m) \equiv M(m,p=0)$ of 
oscillator states as shown
in Fig. 1. It is seen that interactions smooth out the Fermi edge at $m_F=N-1$, but still
leave a gap (not an energy gap!) at $m_F$. 

\begin{figure}[ht]
 \begin{center}
 \epsfig{figure=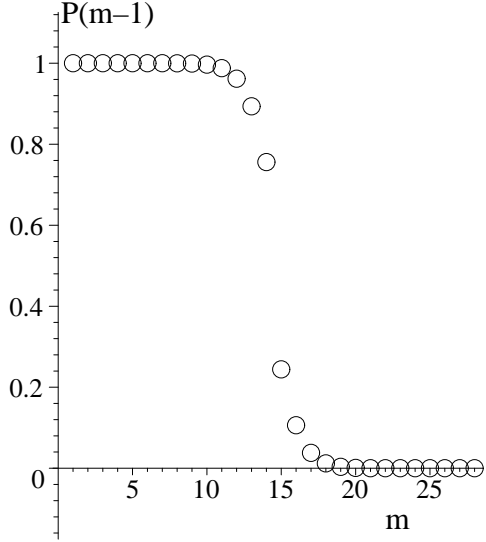,width=0.35\columnwidth}
 \end{center} 
 \caption{\small Occupation probabilities $P$ of oscillator states $m-1$ ($m=1,2,...$)
 for an 
 interacting two-component Fermi gas of $2N=14+14$ atoms in a one-dimensional 
 harmonic trap at zero temperature.  
 Interaction model 1 with $\bar{\alpha}_1=-1$ has been used.
 }
\end{figure}

Fig. 2 displays the off-diagonal matrix elements for $p=1$.
They are significant near the Fermi edge $m_F=N-1$ and cannot be neglected. Their values 
increase further with increasing coupling strength.

\begin{figure}[ht]
 \begin{center}
 \epsfig{figure=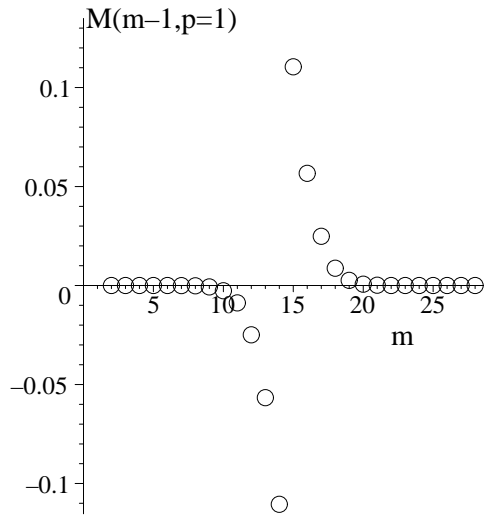,width=0.35\columnwidth}
 \end{center} 
 \caption{\small Off-diagonal matrix elements $M$) versus oscillator state $m-1$ 
 ($m=1,2,...$) for an
 interacting two-component Fermi gas of $2N=14+14$ atoms in a one-dimensional  
 harmonic trap at zero temperature.  
 Interaction model 1 with $\bar{\alpha}_1=-1$ has been used.
 }
\end{figure}

We also present results for the particle density and the momentum 
density. Both are expected to show Friedel oscillations \cite{F58} as noted
in \cite{WW01,GWSZ00}. In agreement with \cite{WW01}, the effective
intra-component interaction, which is always attractive, suppresses the Friedel
oscillations in the particle density

\begin{eqnarray}\label{2.2}
n (z) = \sum _{m = 0}^\infty\sum _{p=-m}^m  \psi _{m-p} 
(z)\,\psi _{m+p} (z)\,M(m,p),
\end{eqnarray} 

as is seen in Fig. 3. In (\ref{2.2}), $\psi _m(z)$ is the oscillator state $|m \rangle $ 
in position representation.

\begin{figure}[ht]			
 \begin{center}
 \epsfig{figure=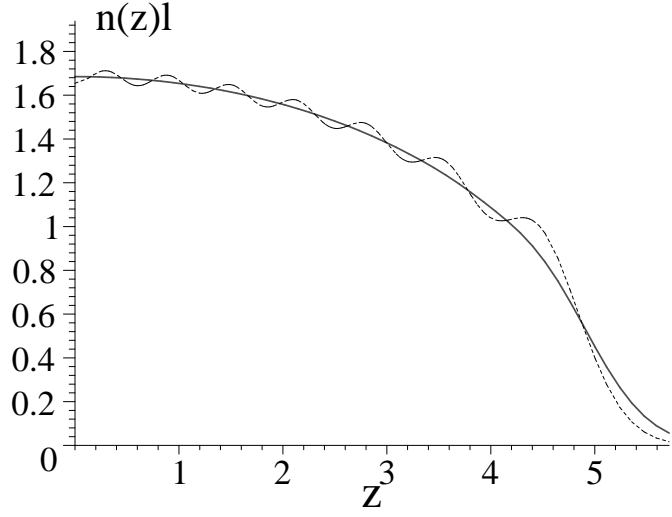,angle=270,width=0.55\columnwidth}
 \end{center} 
 \caption{\small Dimensionless particle density $n(z)\ell$ ($\ell$ is the oscillator length)
 versus dimensionless distance $z$ from the center of the one--dimensional harmonic 
 trap for $2N=14+14$ atoms in the two-component Fermi gas at zero 
 temperature. Broken curve shows unperturbed Friedel oscillations. Smooth 
 curve refers to the interacting case with $\bar{\alpha}_1=-1$. Interaction model 1
 has been used. 
 }
\end{figure}

Conversely, the Friedel oscillations in the momentum density

\begin{eqnarray}\label{2.3}
p(k)=\sum _{m = 0}^\infty\sum _{p=-m}^m (-1)^p \,\psi _{m-p} 
(k)\,\psi _{m+p} (k)\,M(m,p)
\end{eqnarray}

are enhanced \cite{WW01}. This is displayed in Fig. 4 for strong coupling 
($\bar{\alpha}_1=-10$).
We have chosen the oscillator length $\ell \equiv \sqrt{\hbar/(m_A\,\omega _\ell)}$ 
as unit of length, rendering $n(z)$ and $z$ as well as $p(k)$ and $k$ dimensionless.

\begin{figure}[ht]
 \begin{center}
 \epsfig{figure=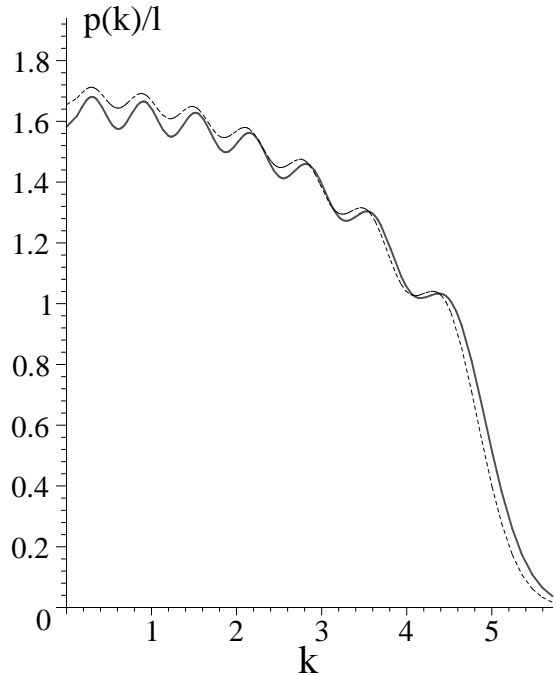,width=0.4\columnwidth}
 \end{center} 
 \caption{\small Dimensionless momentum density $p(k)/\ell$ ($\ell$ is the oscillator length)
 versus dimensionless momentum $k$ for $2N=14+14$ atoms 
 of a two-component Fermi gas in a one-dimensional harmonic at zero 
 temperature. Broken curve shows unperturbed Friedel oscillations. Thick 
 curve refers to the interacting case with $\bar{\alpha}_1=-10$. Interaction model 1
 has been used.  
 }
\end{figure}

In the case of IM2, some modifications occur. We again set $\bar{\alpha}_1
=-1$. We need $\bar{\alpha}_m$ and $\bar{\gamma}_m$ ($m=0,1,2...$) for the evaluation 
of (\ref{1.30}). This requires the knowledge of 
the decay constants $r_\alpha$ and $r _\gamma$ (cf. (\ref{1.20b})). For convenience, 
we set $r_\alpha =r_\gamma \equiv r$ and estimate $r$ by the following argument:
The minimum wave number increment in the trap is $\Delta k \approx 1/L_F \propto
1/\sqrt{N}$, where $L_F=\sqrt{2N-1}$ is the half-width of the classically
allowed
region at the Fermi energy. We, therefore, set $r \approx 1/\sqrt{N}$ or roughly 
$r=0.3$ for the present case $N=14$. This gives $\bar{\alpha}_0=-1.16$ for 
$\bar{\alpha}_1 =-1$ and $\bar{\gamma}_0=1.19$.

Fig. 5 shows the occupation probabilities of oscillator states for IM2. It is seen
that they are more smoothly distributed than in the case of IM1, but still leave a gap at the 
Fermi edge.    

\begin{figure}[ht]
 \begin{center}
 \epsfig{figure=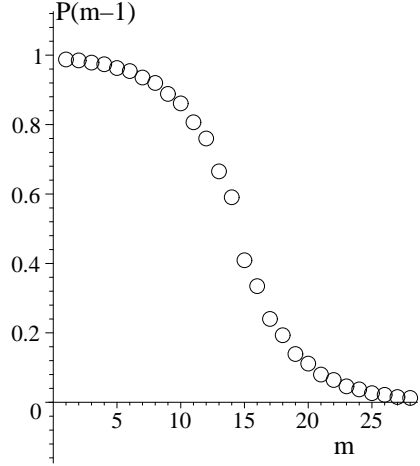,width=0.3\columnwidth}
 \end{center} 
 \caption{\small Occupation probabilities $P$ of oscillator states $m-1$ ($m=1,2,...$)
 for an interacting two-component Fermi gas of $2N=14+14$ atoms in a one-dimensional 
 harmonic trap at zero temperature.  
 Interaction model 2 with $\bar{\alpha}_0=-1.16$ has been used.
 } 
\end{figure}

Finally, we show the momentum density for IM2 in Fig. 6. The Friedel oscillations
are still recognizable for small momenta, but strongly suppressed for momenta
approaching $k_F=\sqrt{2N-1}$.

\begin{figure}[ht]
 \begin{center}
 \epsfig{figure=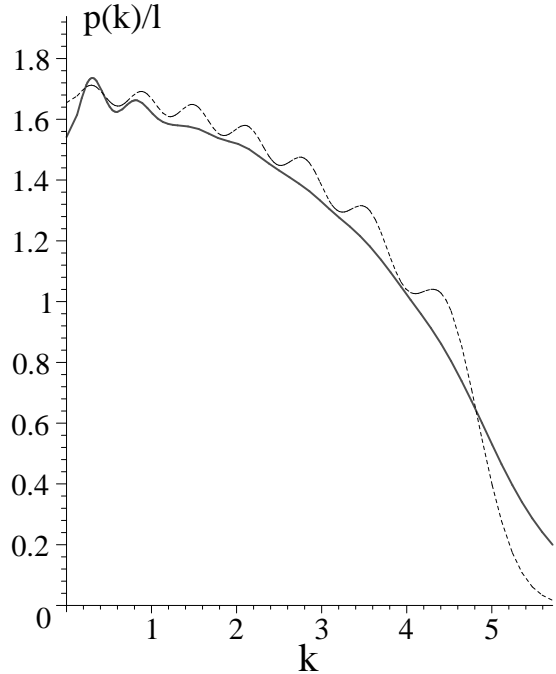,width=0.4\columnwidth}
 \end{center} 
 \caption{\small Dimensionless momentum density $p(k)/\ell$ ($\ell$ is the oscillator length)
 versus dimensionless momentum $k$ for $2N=14+14$ atoms 
 of a two-component Fermi gas in a one-dimensional harmonic at zero 
 temperature. Broken curve shows unperturbed Friedel oscillations. Thick 
 curve refers to the interacting case with $\bar{\alpha}_0=-1.16$. Interaction model 2
 has been used.   
 }
\end{figure}

The off-diagonal matrix elements are significantly smaller for IM2 than for IM1. Nevertheless,
they cannot be neglected: By comparing (\ref{2.2}) with (\ref{2.3}), it is seen that
particle and momentum density would coincide in such an approximation. 

\section{Fermi Edge}

Fig. 1 and also Fig. 5 do not show the gapless distribution of occupation 
probabilities near the Fermi edge $m_F=N-1$, which is characteristic of
a Luttinger liquid, i.e., our system is not a Luttinger liquid. This cannot 
be expected because the system is finite.

We can, however, get a glimpse at Luttinger liquid behaviour in a special limit,
which also presupposes a large particle number $N$.  

First, we consider a very slow decay of the interaction modes $V(m)$ in IM2, i.e., 
$r_\alpha \rightarrow r_\gamma \ll 1$. The factor 

\begin{eqnarray}\label{3.1}
\left[\frac{Z_\gamma}{1+Z_\gamma-\cos(s)}\right]^{\bar{\gamma} _0}\rightarrow
\left(\frac{r_\gamma^2}{r_\gamma^2+s^2}\right)^{\bar{\gamma} _0}
\end{eqnarray}

in the large square brackets 
of the integrand in (\ref{1.30}) then becomes sharply localized at $s=0$.

We now calculate the occupation probability $P(\Delta k_n)$ 

\begin{eqnarray}\label{3.2}
\langle \hat{c}^+ _{N-1+n}\hat{c} _{N-1+n} \rangle =
\langle \hat{c}^+ _{\Delta k_n}\hat{c} _{\Delta k_n} \rangle\equiv P(\Delta k_n) 
\end{eqnarray}

near the Fermi edge and for $N \gg 1$.
$P(\Delta k_n)$ becomes a quasi continuous function of the wave number deviation 
$\Delta k_n =k_n-k_F= n/L_F \rightarrow \Delta k$,
provided  $|n|\ll \mbox{min}(N,1/r_\gamma)$ is fulfilled.  
Using (\ref{3.1}) in (\ref{1.30}) we obtain 

\begin{eqnarray}\label{3.3}
P(\Delta k) =\frac{1}{2}-
\left[_3 F_2\left(\bar{\gamma}_0,\frac{1}{2},1;1,\frac{3}{2};-\left(\frac{\pi}
{r_\gamma}\right)^2 \right)\,r_{\gamma} \,L_F\right] \, \Delta k
\end{eqnarray}

in terms of a generalized hypergeometric function. It is seen that $P(\Delta k)$ 
depends linearly on the wave number deviation in a small region near the Fermi edge.
This can be compared with the Luttinger liquid prediction (cf. e.g., \cite{V95})

\begin{eqnarray}\label{3.4}
P_{LL}(\Delta k) =\frac{1}{2}-\mbox{sgn}\,(\Delta k)\, C \,|\Delta k|^\beta.
\end{eqnarray}

$C$ is a constant and the exponent $\beta$ depends on the Luttinger liquid coupling 
strength $\gamma _{LL}$ according to

\begin{eqnarray}\label{3.5}
\quad \beta =2 \gamma _{LL}\quad \mbox{for}\,\, \gamma _{LL} < \frac{1}{2}, 
\end{eqnarray}
 
and

\begin{eqnarray}
\beta=1 \quad \mbox{for}\,\, \gamma _{LL} \ge \frac{1}{2}.
\end{eqnarray}

We conclude that the above limit of our model agrees with the case 
$\gamma _{LL} \ge 1/2$ of the Luttinger liquid.

\section{Discussion and Summary}

For the interaction to become significant in the quantities calculated, its strength $V(1)$ 
should be as large as $|V(1)| \stackrel{<}{\approx} 0.5 $. We demonstrate that this condition
is within experimental reach. 

To this order, we consider the dipole-dipole interaction \cite{GER01}. It is marginally 
long ranged and thus favors forward scattering. In \cite{GW01}, it is shown that the 
inter-component interaction between longitudinally aligned dipoles reduces exactly to the 
effective one-dimensional potential

$$\tilde{V}_{\rm 1D}(k)=-\frac{\mu _0 \mu^2 \alpha _t^2}{2 \pi}
\left[1-\frac{k^2}{2 \alpha _t^2} \,\exp\left(\frac{k^2}{2 \alpha _t^2}\right)\,
{\rm Ei}\left(-\frac{k^2}{2 \alpha _t^2}\right)\right]$$

in momentum space. Here, $\alpha _t$ is the inverse of the transverse oscillator length, 
$\mu$ the magnetic dipole moment, and ${\rm Ei}$ denotes the exponential integral.

Using this equation in the exact formula (A.13) in \cite{WW01}, $V(1)$ for $N=14$ 
is found to be

$$V(1) = 0.8 \,\left(\frac{\mu _0 \mu^2 m_A^{3/2} \omega _\ell^{1/2}}{2 \pi \hbar^{5/2}}
\right)\,\frac{1}{F}.$$

The quantity $F$ denotes the filling factor $F=N \omega _\ell/\omega _t$. For example in
$^{53}{\rm Cr}$, $V(1)$ becomes of the required magnitude provided $F$ is very small,
i.e., the trap is highly anisotropic.

In summary, the bosonization method has been used to construct a theory for a two component 
gas of spin
polarized fermions in a one-dimensional harmonic potential with forward
scattering between the two components. Asymptotic results with respect to the fermion 
number $N$ were obtained for the one-particle matrix elements and used to discuss
occupation probabilities for oscillator states, off-diagonal matrix elements, and
distribution functions for particles and momenta in the harmonic trap.
All these quantities can be significantly affected by the attractive interaction generated 
within each component. Specifically, the Friedel oscillations in the particle density 
are suppressed, while they survive in the momentum density.

It has to be seen, whether the predicted Friedel oscillations can be observed experimentally.
The amplitudes of the Friedel oscillations scale as $1/N$ \cite{GWSZ00}, hence Friedel 
oscillations are unobservable in a macroscopic bounded Fermi sea. Small particle numbers
pose, however, severe detection problems. 
A conceivable experimental method to observe Friedel oscillations for atom numbers 
of the order of $100$ is indicated in \cite{GWSZ00}. The method proposes microfabrication 
techniques to produce arrays of microtraps.
		  
On the other hand, the
asymptotic bosonization method requires particle numbers, which are not too small. 
This is due to 
the presence of the anomalous vacuum, which couples to the real particles. For instance,
the sum rule $\sum _n P(n) =N$ gives a somewhat larger value than the number $N$ of real 
particles when (\ref{1.28}) or (\ref{1.30}) are used. The excess $\Delta N >0$ grows with 
coupling strength and decreasing particle number. For $N=14$ and very strong coupling 
$\bar{\alpha} _1 =-10$, $\Delta N$ is about $8 \cdot 10^{-3}$, for $\bar{\alpha} _1=-1$, 
$\Delta N$ is less than $10^{-10}$.  

The atom number $2N=14+14$ and the coupling values employed here are appropriate to give 
visible Friedel oscillations and reliable results of the bosonization method.

{\bf Acknowledgements}: The authors thank S. N. Artemenko, F. Gleisberg, and W. P. Schleich
for valuable discussions and the Deutsche Forschungsgemeinschaft for financial support.

\section{Appendix}

\newcounter{affix}
\setcounter{equation}{0}
\setcounter{affix}{1}
\renewcommand{\theequation}{\Alph{affix}.\arabic{equation}}

In this Appendix, we extend the bosonization procedure in \cite{SchM96} to the case of two 
components.

Instead of $\hat{d}_{p \pm}$ operators, which are needed for the diagonalization of the 
interacting Hamiltonian, the following set of operators play the role of
the $\hat{b}$ and $\hat{b}^+$ operators ($n \ge 1$) in \cite{SchM96}:

\begin{eqnarray}\label{A.1}
\hat{b}^+_{n \sigma}\equiv \frac{1}{\sqrt{2}}\,[\hat{d}^+_{n +}+\sigma \hat{d}^+_{n -} ],
\quad\hat{b}_{n \sigma}\equiv \frac{1}{\sqrt{2}}\,[\hat{d}_{n +}+\sigma \hat{d}_{n -} ].
\end{eqnarray}

They are canonical conjugates. Evidently:

\begin{eqnarray}\label{A.2}
\hat{b}^+_{n \sigma}\equiv \frac{1}{\sqrt{n}}\,\hat{\rho}_\sigma(n). 
\end{eqnarray}

Then the two relations hold

\begin{eqnarray}\label{A.3}
[\hat{b}^+_{n \sigma},\hat{c}^+_{k \sigma'}\hat{c}_{l \sigma'}]
&=&\delta _{\sigma,\sigma'}\,\frac{1}{\sqrt{n}}\,(\hat{c}^+_{k+n \sigma}\hat{c}_{l \sigma}
-\hat{c}^+_{k \sigma}\hat{c}_{l-n \sigma}),
\\[4mm]\nonumber
[\hat{b}_{n \sigma},\hat{c}^+_{k \sigma'}\hat{c}_{l \sigma'}]
&=&\delta _{\sigma,\sigma'}\,\frac{1}{\sqrt{n}}\,(\hat{c}^+_{k-n \sigma}\hat{c}_{l \sigma}
-\hat{c}^+_{k \sigma}\hat{c}_{l+n \sigma}).
\end{eqnarray}

Following the arguments in \cite{SchM96}, the associated Bose fields for $\sigma=\sigma'$ are

\begin{eqnarray}\label{A.4}
\hat{\phi}_\sigma (v) = -i \sum^\infty _{n = 1} \frac{1}{ \sqrt {n} }\, e^{i n v}
\,\hat{b}_{n \sigma} 
\equiv -i \sum^\infty _{n = 1} \frac{1}{ \sqrt {2 n} }\, e^{i n v}
\,(\hat{d}_{n+} +\sigma \hat{d}_{n-})\neq \hat{\phi}_\sigma^+(v).
\end{eqnarray}


\begin{thebibliography}{111111}

\bibitem{ADB95}
M. H. Anderson {\it et al.}, Science {\bf 269}, 198 (1995); K. B. Davis {\it et al.}, 
Phys. Rev. Lett. {\bf 75}, 3969 (1995); C. C. Bradley {\it et al.}, {\it ibid.} {\bf 75},
1687 (1995).

\bibitem{BDL98}
F. Brosens, J. T. Devreese, and L. F. Lemmens, Phys. Rev. E {\bf 57}, 3871 (1998).

\bibitem{BB98}
G. M. Bruun and K. Burnett, Phys. Rev. A {\bf 58}, 2427 (1998).

\bibitem{ZGOM00} M. A. Zaluska-Kotur, M. Gajda, A. Orlowski, and J. Mostowski,
Phys. Rev. A {\bf 61}, 033613 (2000).

\bibitem{HFSt97} 
M. Houbiers, R. Ferwerda, H. T. C. Stoof, W. I. McAlexander, C. A. Sackett and R. G. Hulet, 
Phys. Rev. A {\bf 56}, 4864 (1997).

\bibitem{BP98}
M. A. Baranov and D. S. Petrov, Phys. Rev. A {\bf 58}, R801 (1998).

\bibitem{HSt99}
M. Houbiers and H. T. C. Stoof, Phys. Rev. A {\bf 59}, 1556 (1999).

\bibitem{C99}
R. Combescot, Phys. Rev. Lett. {\bf 83}, 3766 (1999).

\bibitem{DMJ99}
B. DeMarco and D. S. Jin, Science {\bf 285}, 1703 (1999).

\bibitem{SFCCKMS01}
F. Schreck, G. Ferrari, K. L. Corwin, J. Cuizolles, L. Khaykovich, M.-O. Mewes,
and C. Salomon, Phys. Rev. A {\bf 64}, 011402(R) (2001).

\bibitem{VFP98} V. Vuletic, T. Fischer, M. Praeger, T. W. H\"ansch, 
and C. Zimmermann, Phys. Rev. Lett. {\bf 80}, 1634 (1998).

\bibitem{FGZ98}
J. Fortagh, A. Grossmann, C. Zimmermann, and T. W. H\"ansch,
Phys. Rev. Lett. {\bf 81}, 5310 (1998).

\bibitem{DCS99}
J. Denschlag, D. Cassettari, and J. Schmiedmayer, Phys. Rev. Lett.
{\bf 82}, 2014 (1999).

\bibitem{RHH99}
J. Reichel, W. H\"ansel, and T. W. H\"ansch, Phys. Rev. Lett.
{\bf 83}, 3398 (1999).

\bibitem{GER01}
K. Goral, B-G. Englert, and K. Rzazewski, Phys. Rev. A {\bf 63}, 033606 (2001).

\bibitem{BBC00}
H. L. Bethlem, G. Berden, F. M. H. Crompvoets, R. T. Jongma, A. J. A. van Roij, 
and G. Meijer, Nature (London) {\bf 406}, 491 (2000).

\bibitem{C84}
J. L. Cardy, J. Phys. A {\bf 17}, L385 (1984).

\bibitem{EA92}
S. Eggert and I. Affleck, Phys. Rev. B {\bf 46}, 10866 (1992).

\bibitem{FG95} 
M. Fabrizio and A. O. Gogolin, Phys. Rev. B {\bf 51}, 17827 (1995).

\bibitem{EG95}
R. Egger and H. Grabert, Phys. Rev. Lett. {\bf 75}, 3505 (1995).

\bibitem{WVF96}
Y. Wang, J. Voit, and Fu-Cho Pu, Phys. Rev. B {\bf 54}, 8491 (1996).

\bibitem{MEJ97}
A. E. Mattsson, S. Eggert, and H. Johannesson, Phys. Rev. B {\bf 56}, 15615 (1997).

\bibitem{VYG00}
J. Voit, Yupeng Wang, and M. Grioni, Phys. Rev. B {\bf 61}, 7930 (2000).

\bibitem{WW01}
W. Wonneberger, Phys. Rev. A {\bf 63}, 063607 (2001).

\bibitem{E79}
V. J. Emery, {\it Theory of the One-Dimensional Electron Gas}, in {\it Highly Conducting
One-Dimensional Solids}, edited by J. T. Devreese, R. P. Evard, and V. E. van Doren 
(Plenum, New York, 1979), p247.

\bibitem{V95}
J. Voit, Rep. Prog. Phys. {\bf 58}, 977 (1995).

\bibitem{Sch95}
H. J. Schulz {\it Fermi Liquids and Non-Fermi Liquids}, in {\it Mesoscopic Quantum Physics},
edited by E. Akkermans, G. Montambaux, J.-L. Pichard, and J. Zinn-Justin (Elsevier, 
Amsterdam, 1995), p533.

\bibitem{LE74}
A. Luther and V. J. Emery, Phys. Rev. Lett. {\bf 33}, 589 (1974).


\bibitem{DL73}
I. E. Dzyaloshinskii and A. I. Larkin, Zh. Eksp. Teor. Fiz. {\bf 65}, 411 (1973);
(Sov. Phys. JETP {\bf 38}, 202 (1974)).

\bibitem{F76}
H. C. Fogedby, J. Phys. C: Solid State Phys. {\bf 9}, 3757 (1976).

\bibitem{GMR79}
G. Grinstein, P. Minnhagen, and A. Rosengren, J. Phys. C: Solid State Phys.
{\bf 12}, 1271 (1979).

\bibitem{Haldane}
F. D. M. Haldane, J. Phys. C: Solid State Phys. {\bf 14}, 2585 (1981).

\bibitem{VMT00}
P. Vignolo, A. Minguzzi, and M. P. Tosi, Phys. Rev. Lett. {\bf 85}, 2850 (2000).

\bibitem{GWSZ00}
F. Gleisberg, W. Wonneberger, U. Schl\"oder, and C. Zimmermann, Phys. Rev. A {\bf 62},
063602 (2000).

\bibitem{GW01}
F. Gleisberg and W. Wonneberger, unpublished.

\bibitem{deK35}
R. de L. Kronig, Physica {\bf 2}, 968 (1935).

\bibitem{LP74}
A. Luther and I. Peschel, Phys. Rev. B {\bf 9}, 2911 (1974).

\bibitem{M74}
D. C. Mattis, J. Math. Phys. {\bf 15}, 609 (1974).

\bibitem{HSU80}
R. Heidenreich, R. Seiler, and D.A. Uhlenbrock, J. Stat. Phys. {\bf 22}, 27 (1980).

\bibitem{SchM96}
K. Sch\"onhammer and V. Meden, Am. J. Phys. {\bf 64}, 1168 (1996).

\bibitem{F58} J. Friedel, Nuovo Cimento Suppl. {\bf 7}, 287 (1958).





\end{thebibliography}
\end{document}